# Machine learning-enabled atomistic insights into phase boundary engineering of solid-solution ferroelectrics

Weiru Wen[1, 2], Fan-Da Zeng[2], Ben Xu[3], Bi Ke[1, *], Zhipeng Xing[2, *], Hao-Cheng Thong[2, *], Ke Wang[2]

[1] State Key Laboratory of Information Photonics and Optical Communications, School of Physical Science and Technology, Beijing University of Posts and Telecommunications, Beijing, 100876, China

[2] State Key Laboratory of New Ceramic Materials, School of Materials Science and Engineering, Tsinghua University, Beijing 100084, P. R. China

[3] Graduate School, China Academy of Engineering Physics, Beijing 100193, People's Republic of China

* Corresponding authors: Bi Ke (bike@bupt.edu.cn); Zhipeng Xing (xingzhipeng@tsinghua.edu.cn); Hao-Cheng Thong (haocheng@mail.tsinghua.edu.cn)

Abstract

Atomistic control of phase boundaries is crucial for optimizing the functional properties of solid-solution ferroelectrics, yet their microstructural mechanisms remain elusive. Here, we harness machine-learning-driven molecular dynamics to resolve the phase boundary behavior in the $KNbO_3$–$KTaO_3$ (KNTO) system. Our simulations reveal that chemical composition and ordering enable precise modulation of polymorphic phase boundaries (PPBs), offering a versatile pathway for materials engineering. Diffused PPBs and polar nano regions, predicted by our model, highly match with experiments, underscoring the fidelity of the machine-learning atomistic simulation. Crucially, we identify elastic and electrostatic mismatches between ferroelectric $KNbO_3$ and paraelectric $KTaO_3$ as the driving forces behind complex microstructural evolution. This work not only resolves the longstanding microstructural debate but also establishes a generalizable framework for phase boundary engineering toward next-generation high-performance ferroelectrics.

# 1. Introduction

Ferroelectrics are versatile functional materials that are extensively applied in the fields of energy storage devices, memory devices, actuators, and sensors (*1-6*). Phase boundary engineering, including the composition-dependent morphotropic phase boundary (MPB) (*7*) and the temperature-dependent polymorphic phase boundary (PPB) (*8*), has been a useful strategy for enhancing the properties of ferroelectrics. Phase boundary engineering is commonly achieved by forming solid solutions. For instance, the solid solution of antiferroelectric $PbZrO_3$ and ferroelectric $PbTiO_3$ results in a classic MPB in the $Pb(Zr, Ti)O_3$ (PZT) system, leading to optimal electromechanical properties (*9, 10*). Similarly, the solid solutions of relaxors and ferroelectrics, like (Bi, Na)$TiO_3$-$BaTiO_3$ and Pb(Mg, Nb)$O_3$-$PbTiO_3$ systems, can also realize MPBs that highly amplify the electro-strain for actuating applications (*11-13*). Construction of ferroelectric-paraelectric PPBs near room temperature in $BaTiO_3$-based solid solution can promote the electrocaloric effect significantly (*14, 15*). In another instance, the thermal stability of the prominent lead-free piezoelectric ceramic (K, Na)$NbO_3$ is notably enhanced through the construction of a diffused PPB (*16, 17*). Despite the widespread adoption of phase boundary engineering, the underlying mechanism is still highly debatable, *e.g.*, the MPB in PZT was historically attributed to the coexistence of rhombohedral and tetragonal phases (*18*), which was later challenged by the discovery of a monoclinic phase acting as a bridging structure (*19*).

Theoretical atomistic simulation can be a promising avenue for resolving the sophisticated microstructure at phase boundaries (*20, 21*). Still, the simulations of solid solutions are often limited by the computational resources, as a reasonable accountability for concentration and chemical distribution requires an enormous modelling system, *e.g.*, a 1% substitution concentration necessitates at least several hundred atoms, even without considering the distribution. Unfortunately, first-principles calculations like density functional theory (DFT), while capable of capturing intricate electron-nuclear interactions, are computationally prohibitive for these large-scale simulations, limiting their application in both spatial and temporal dimensions (*22*). In contrast, molecular dynamics (MD) simulations can handle larger systems but often lack the interatomic potentials of sufficient accuracy (*22*). Recent advances in machine learning have enabled a better fitting of the potential energy surface, *e.g.*, using Gaussian approximation (*23, 24*), graph neural network (*25-27*), and feed-forward neural network (*28, 29*). The machine-learning interatomic potentials have been proven powerful for simulating a wide range of ferroelectrics (*30-33*), including conventional ferroelectric perovskites, hafnium oxides, and van der Waals ferroelectrics, and have been recently adopted for solid solutions. The improved predictive power of MD simulations based on machine-learning interatomic potentials should significantly deepen our microstructural understanding of ferroelectrics, particularly at complex phase boundaries.

Among these materials, K(Nb, Ta)$O_3$ (KNTO) is a particularly versatile solid solution system for electro-optical and electromechanical applications (*34-36*), formed by mixing ferroelectric $KNbO_3$ (KNO) and paraelectric $KTaO_3$ (KTO). KNTO compositions exhibit enhanced quadratic electro-optic coefficients near the paraelectric-ferroelectric PPB (*37, 38*). Specifically, compositions with Nb/Ta ratios around 3:7 display relaxor behavior and superior performance, believed to be driven by polar nanoregions (PNRs) embedded within a non-polar matrix (*39-42*). Furthermore, tuning the Nb/Ta ratio allows the ferroelectric-ferroelectric PPB to be shifted to room temperature, enhancing dielectric susceptibility and electromechanical coupling (*43*). A comprehensive atomistic understanding of the PPB in KNTO, as a function of temperature and composition, is of significant scientific interest. However, theoretical studies on this system remain limited, with most previous work focusing on small-scale configurations, such as KNO-KTO layered superlattices (*44*). Recently, following our previous work on KNO (*45*), Xing *et al.* investigated the KNTO solid solution system using machine-learning MD (*46*), focusing on the phonon analysis of a few specific compositions and compared with the neutron scattering experiment. The study magnificently revealed the variation of the order-disorder nature upon Ta substitution into KNO, yet barely focus on the variation of the PPB, which is the key component in materials engineering.

In this study, we develop a deep-learning interatomic potential (or deep potential, DP) model tailored for KNTO solid solutions to investigate PPB behavior across temperature and composition. Utilizing the DP model, we can qualitatively reproduce the experimentally observed temperature-composition phase diagram. Macroscopic phase structures are identified via statistical calculation of local atomic displacements. Different types of PPBs are clearly distinguished and shown to be tunable through composition and chemical ordering. Interestingly, in KNTO compositions with high KTO content, the paraelectric-ferroelectric PPB exhibits diffused and relaxor-like characteristics, potentially explaining the experimentally observed enhancements in electro-optical performance. These phenomena are further interpreted through analyses of elastic and electrostatic mismatches within lattices. Complementary experimental characterizations are also presented to support and validate the theoretical findings.

## 2. Results and discussion

2.1 DP model

The development of the DP model is achieved through a concurrent learning framework facilitated by the DPGEN module (Figure 1a). The framework involved iterative short model training (DeePMD), exploration via MD simulations (LAMMPS), and first-principles DFT data labeling (VASP), where details can be found in the methods section. The training dataset for the KNTO solid solution was constructed by augmenting the established KNO dataset (13635 configurations) (*45*) with configurations containing 25 mol%, 50 mol%, 75 mol%, and 100 mol% KTO. Through an effective exploration strategy over varying compositions, phases (including cubic *Pm*-3*m*, tetragonal *P*4*mm*, orthorhombic *Amm*2, and rhombohedral *R*3*m*), and thermodynamic conditions (including temperatures and pressures), the final dataset encompasses 15282 configurations, with added KNTO configurations accounting for ~11% of the original KNO dataset (Figure 1b). In later sections, we will show that adding such a small amount of data is capable of describing the whole solid-solution system. The parameterization of the first-principles Born-Oppenheimer potential energy surface for a more efficient atomic simulation is also named the second-principles model (*22*).

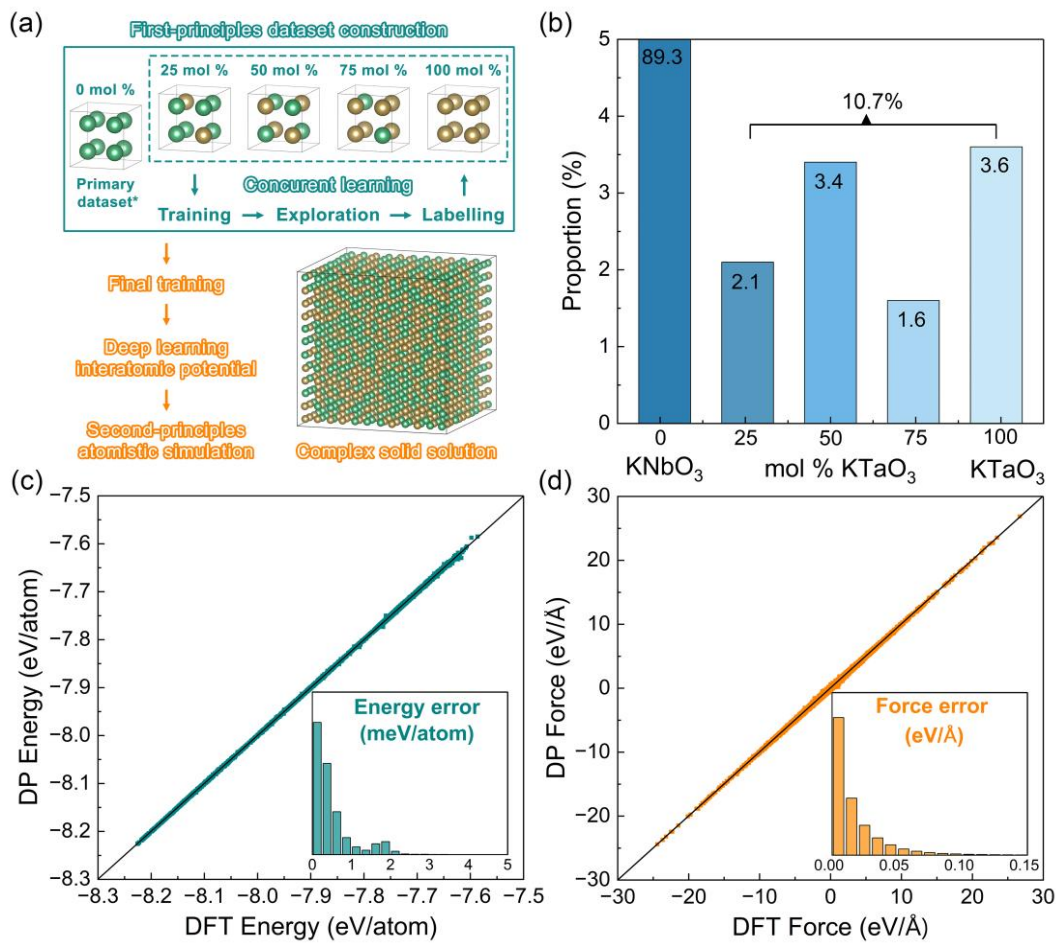

Figure 1. Development and benchmark of the DP model. (a) Construction workflow of the DP model for the KNTO solid solution. (b) Proportion of configurations with different chemical compositions in the dataset. Validation of the DP model on (c) potential energy and (d) atomic force towards all configurations in the dataset. The insets show the distributions of absolute errors.

Figures 1(c)-(d) show the benchmarking of the final DP model on the whole dataset in terms of potential energy and atomic forces, with the errors at approximately $10^{-3}$~$10^{-4}$ eV/atom and ~$10^{-2}$ eV/Å, respectively. The benchmarking results indicate that the DP model has perfectly learned the training dataset and possibly has a good capability for simulating the KNTO solid solution. The high accuracy of the DP model

can ensure precise discerning of subtle energy differences among different phase structures of KNTO, highlighting its universal modeling capability for complex solid-solution systems.

2.2 Temperature-dependent phase diagram

MD simulation based on the DP model is employed to simulate the temperature-dependent phase diagram of the KNTO solid solution, as illustrated in Figure 2(a). First, the paraelectric-to-ferroelectric phase transition temperature (*i.e.*, cubic phase to tetragonal phase) almost linearly decreases as a function of KTO concentration. The tetragonal-to-orthorhombic phase transition temperature exhibits a similar linear variation, though with a smaller slope. On the contrary, the orthorhombic-to-rhombohedral phase transition temperature rises with increasing KTO concentration until ~50 mol % and is subsequently followed by a significant decrease. The DP-simulated phase diagram shows strong similarity to the previously reported experimental data (*47*), as shown in Figure 2(b).

To understand the variation, additional simulations are conducted at 0 K for KNTO with different KTO concentrations. Potential energy differences of various phases relative to the ground-state rhombohedral phase are computed, as shown in Figure 2(c). In most cases, the energy difference decreases with increasing KTO concentration, except for the energy difference between orthorhombic and rhombohedral phases, which first increases and then decreases, with a turning point near 50 mol%. Fundamentally, the total energy comprises both potential and kinetic contributions. While kinetic energy is directly proportional to temperature, potential energy varies across systems. Therefore, one might expect a linear correlation between the phase transition temperature and the potential energy difference between each phase and the ground state. Figure 2(d) plots the phase transition temperatures as a function of the calculated potential energy differences. All systems exhibit a positive correlation, yet non-linear. This non-linear deviation is likely attributable to simulation parameters such as the choice of thermostat, barostat, cooling rate, and chemical distribution. A systematic investigation of these factors is required in a further study to achieve more precise quantitative predictions. Nevertheless, the present study focuses on the qualitative influence of KTO content on the phase behavior.

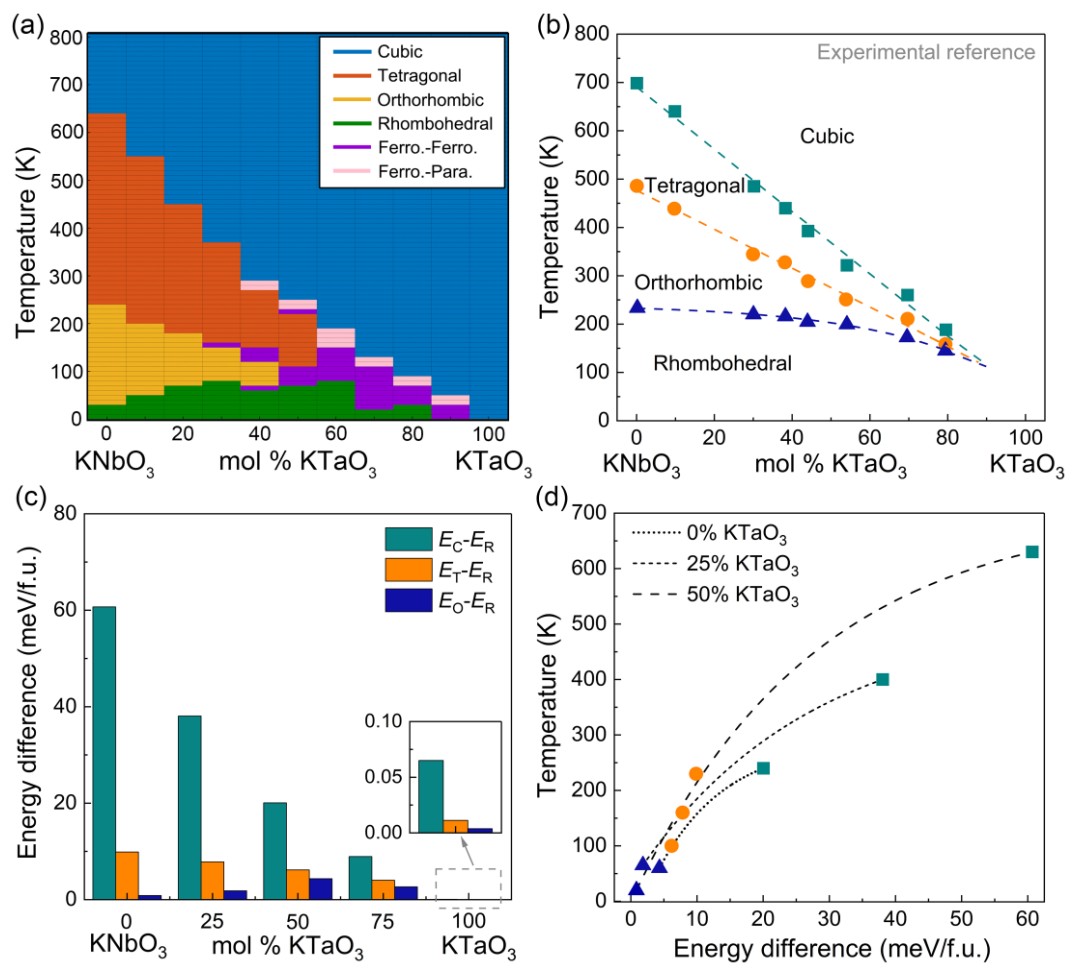

Figure 2. Phase diagram of KNTO. (a) DP-simulated and (b) experimentally reported temperature-dependent phase diagram of KNTO (*47*). (c) Energy differences between phases (including cubic, tetragonal, and orthorhombic) and the ground-state rhombohedral phase. (d) The relationship between the phase transition temperature captured from the DP-simulated phase diagram and the energy difference among phases. Please note that the phase transition temperatures of KNTOs with 25 mol % are determined by averaging from those of the neighboring KNTOs (20 mol % and 30 mol %), while the KNTOs with 75 mol % and 100 mol % are not presented due to the difficulty of defining the phase transition temperature.

In addition to the variation in phase transition temperatures, the phase structures become increasingly complex and ambiguous with higher KTO concentrations. The simulated phase diagram can be segmented into several distinct regions based on KTO content. When the KTO concentration is below 30 mol %, systems undergo typical sequential phase transitions from cubic to tetragonal, orthorhombic, and rhombohedral structures upon cooling. However, as the KTO concentration increases beyond this threshold, the phase transitions become increasingly diffused, with multiple phases coexisting over a broader temperature range near the transition points. The diffused phase transition regions, known as the PPB aforementioned, are marked with pink and purple colors in Figure 2(a), referring to the paraelectric-ferroelectric (denoted as Para-Ferro) and ferroelectric-ferroelectric (denoted as Ferro-Ferro) phase coexistence, respectively. Notably, for KTO concentrations exceeding 50 mol %, the orthorhombic phase region is effectively suppressed, replaced by a wider ferroelectric-ferroelectric PPB. The increasing diffuseness of this ferroelectric-ferroelectric PPB is particularly advantageous for achieving temperature-insensitive properties (*48, 49*).

The evolution of phase structures is further detailed in Figure 3, which presents the frequency of observed phases as a function of temperature for selected compositions, including 20 mol %, 40 mol %, 60 mol %, and 80 mol % KTO. The pure end members (KNO and KTO) are excluded from this analysis due to their relatively simple phase behavior. Phase classification is performed based on the local atomic displacements using the algorithm described in the methods section. As shown in Figure 3, sharp and well-defined phase transitions are observed in the sample with 20 mol % KTO. In contrast, compositions with higher KTO concentrations exhibit significantly broader transition ranges, clearly capturing the diffused nature of the phase transition. Specifically, in the 60 mol % and 80 mol % KTO samples, both ferroelectric and paraelectric phases coexist at the PPB. Notably, these compositions have been experimentally reported to exhibit relaxor behavior (*50, 51*).

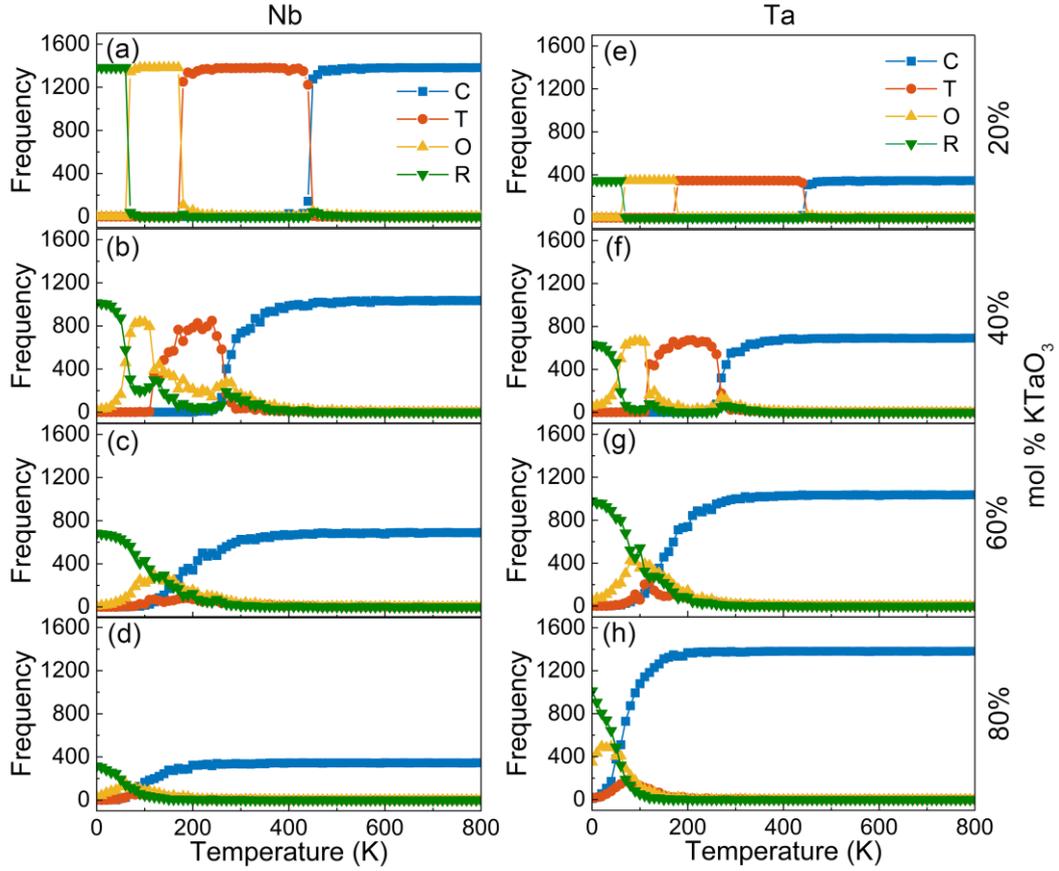

Figure 3. Variation of phase structure at local unit cells in KNTOs with different KTO concentrations. Statistics of the phase structure at KNO unit cells in KNTOs with (a) 20 mol %, (b) 40 mol %, (c) 60 mol %, and (d) 80 mol % KTO. Statistics of the phase structure at KTO unit cells in KNTOs with (e) 20 mol %, (f) 40 mol %, (g) 60 mol %, and (h) 80 mol % KTO. C, T, O, and R represent cubic, tetragonal, orthorhombic, and rhombohedral phases, respectively.

2.3 Diffuseness of PPBs

The diffuseness of PPBs is usually pronounced in real ceramic samples. As shown in Figure 4(a)-(b), temperature-dependent dielectric measurements were performed on the K(Nb$_{1-x}$Ta$_x$)O$_3$ ($x$ = 0, 0.1, 0.2, 0.3) ceramic samples at a frequency of 100 kHz. The results demonstrate that the diffuseness of PPBs increases with the KTO concentration. The PPBs in KNTO with 30 mol % KTO spread over several tens of K, which strongly deviates from the simulation. This discrepancy may originate from chemical inhomogeneity in the samples, which could amplify the diffuseness. In the simulations, a uniform chemical distribution was assumed, whereas electron probe microanalysis (EPMA) observations, as illustrated in panels 4(c)-(e), demonstrate a non-uniform spatial distribution of Ta and Nb elements. Detailed comparison of BSE signals from scanning electron microscopy (SEM) among all samples can be found in Figure S2.

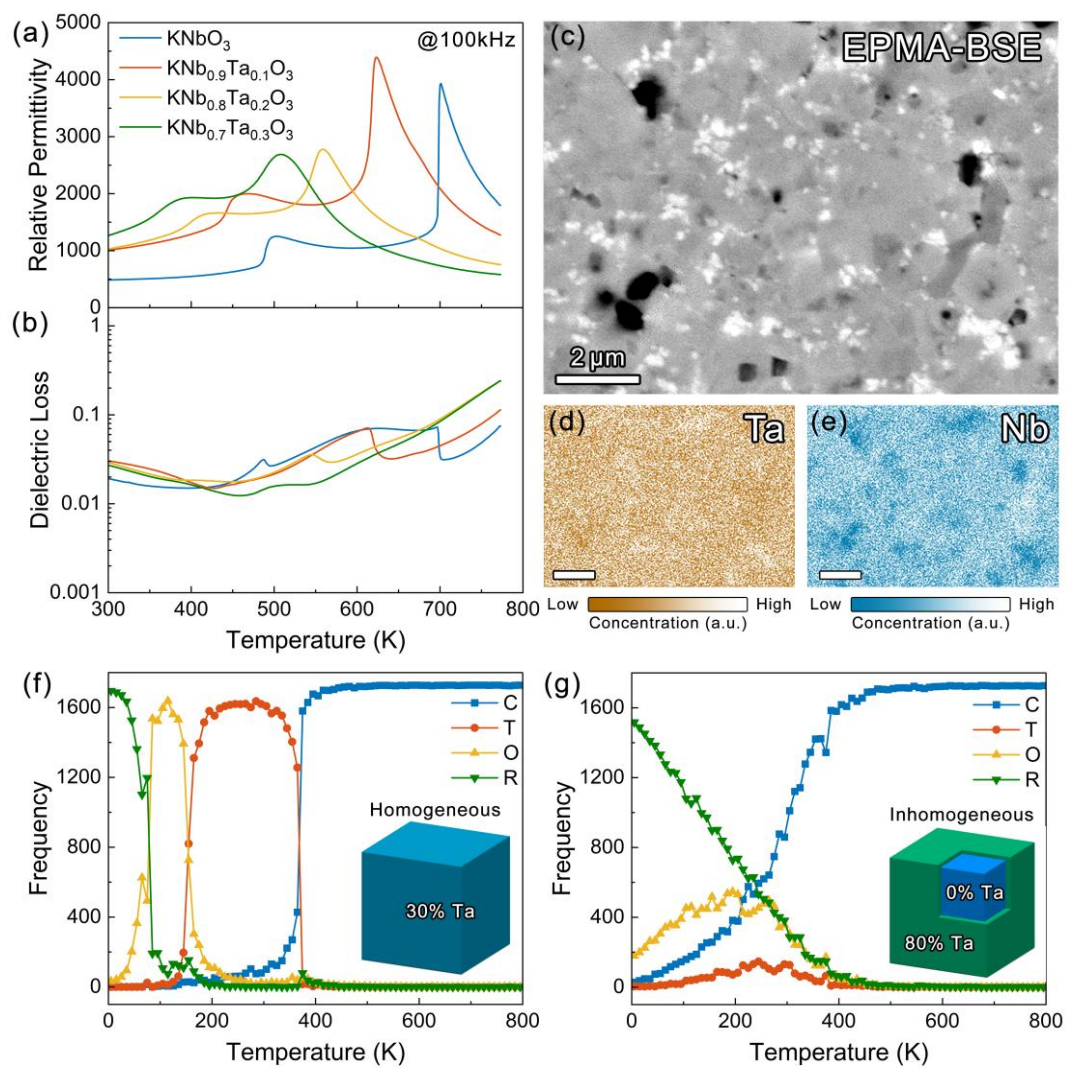

Figure 4. Influence of chemical inhomogeneity on the phase transition of KNTO. Temperature-dependent (a) relative permittivity and (b) dielectric loss of $K(Nb_{1-x}Ta_x)O_3$ ($x$ = 0, 0.1, 0.2, 0.3) ceramics. (c) EPMA-BSE signal of the polished surface of $K(Nb_{0.7}Ta_{0.3})O_3$. Distribution of (d) Ta and (e) Nb probed by EPMA-WDS signals. Simulation of the temperature-dependent phase transitions of chemically (f) homogeneous and (g) inhomogeneous $K(Nb_{0.7}Ta_{0.3})O_3$ supercells.

To further validate the influence of chemical inhomogeneity on diffuseness observed in experiments, an inhomogeneous core-shell structure is constructed as illustrated in Figure 4(g), and systematically compared with the homogeneous distribution case presented in Figure 4(f). The inhomogeneous model is a 12×12×12 supercell comprising a 10×10×10 KNO core and a 2-layer KNTO shell with 80% Ta concentration, maintaining an equivalent Ta concentration of 30% across the entire system. A more significant diffused phase transition is observed in the temperature-dependent phase transition simulation of the inhomogeneous model. The result further corroborates that chemical ordering is decisive to the diffuseness of the PPBs. With proper control of processing conditions, different chemical ordering can be achieved (*52, 53*), thereby providing flexibility to control the diffuseness of PPBs for desired functional properties.

2.4 Interplay between KNO and KTO

Fundamentally, KNO is ferroelectric, whereas KTO is paraelectric. According to soft-mode theory, the spontaneous polarization in ferroelectrics originates from ferroelectric instability, which is characterized by the softening of a specific transverse optical phonon mode at the Γ point (*54*). This soft mode can be revealed by calculating the phonon spectrum of the highest-symmetry phase, namely the cubic phase, as illustrated in Figure S1. The phonon spectra reveal the presence of an imaginary mode at the Γ point in both KNO and KTO, indicative of ferroelectric instability. However, the imaginary frequency in KNO is significantly larger than that in KTO, implying a much stronger ferroelectric instability in KNO. In contrast, the instability in KTO is extremely weak and is typically suppressed by quantum fluctuations, thus termed as quantum paraelectric. Even if the ferroelectricity can be observed, the resulting polarization is expected to be minimal due to the inherent weakness of the instability.

In a solid solution composed of ferroelectric KNO and paraelectric KTO, elastic and electrostatic mismatches are inevitable due to differences in lattice parameters and polarization magnitude. Figures 5(a) and 5(b) illustrate supercells composed of tetragonal unit cells, where 10 KNO units are combined with 5, 10, or 15 KTO units. The polarization orientations are intentionally set to be parallel or perpendicular to the KNO-KTO interface in two different cases, respectively. It is important to note that strain and polarization are inherently coupled within ferroelectric materials, and these two effects cannot be entirely decoupled in such simulations. Nonetheless, the following analyses try to isolate and highlight the distinct roles played by elastic and electrostatic mismatches.

In the parallel configuration (Figure 5(a)), KNO cells exhibit significantly stronger polarization than KTO cells, and the overall polarization of the system decreases as the number of KTO units increases. The induced polarization in KTO originates from the elongated cells due to the tensile strain from KNO cells, while the reduced polarization in KNO originates from the compressed cells due to the compressive strain from KTO cells. This simulation highlights the influence of elastic mismatch. In contrast, when the polarization is oriented perpendicular to the interface (Figure 5(b)), a gradual reduction of polarization magnitude is observed in KNO approaching the interface, while KTO cells exhibit a progressive increase. This trend suggests the presence of an interfacial electrostatic field arising from polarization mismatch between KNO and KTO. It is interesting to note that while lattice mismatch can be accommodated by homogeneously distributing strain across the entire system, polarization mismatch exerts a localized effect, primarily influencing the regions near the interfaces.

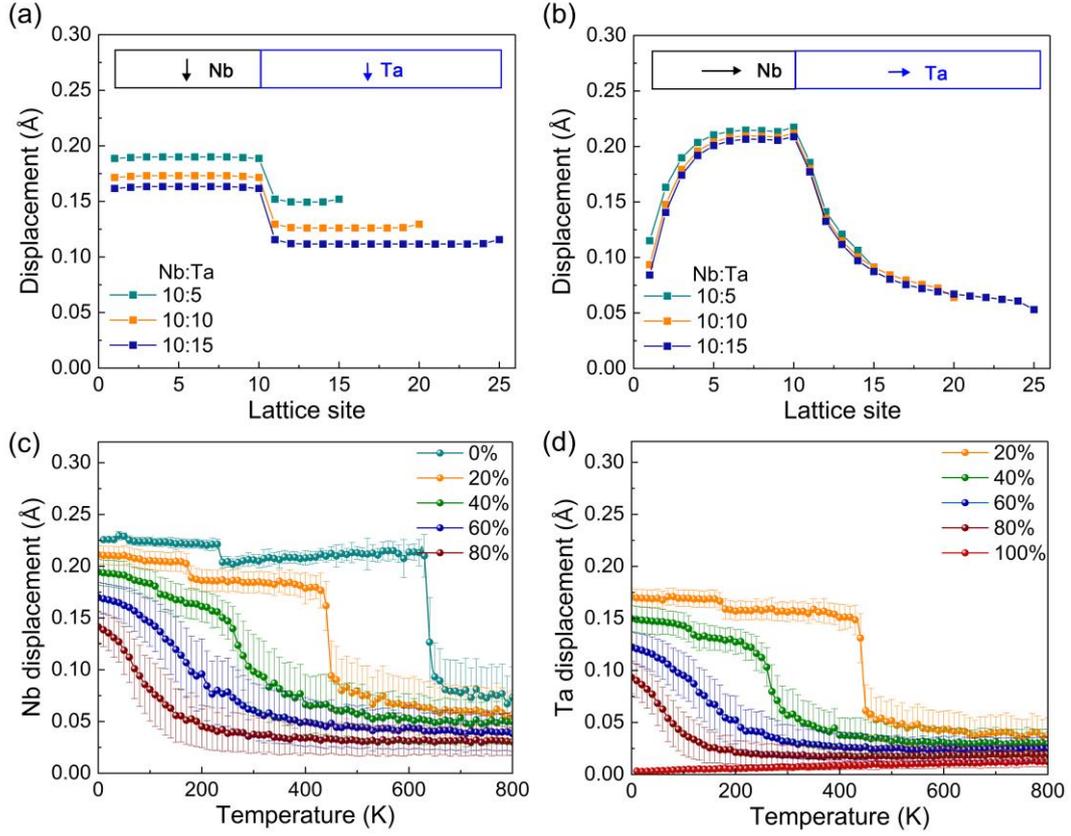

Figure 5. The effective polarization of KNO and KTO cells when combined with polarization direction (a) parallel with and (b) perpendicular to the interface. The effective polarizations of (c) KNO and (d) KTO cells in KNTOs with different KTO concentrations. The data shows the statistical average over the whole space with a standard deviation.

Figures 5(c) and 5(d) present the polarization magnitudes of KNO and KTO components in KNTO during the temperature-dependent simulations shown in Figure 2(a). As expected, the overall polarization magnitude decreases with increasing KTO concentration, reflecting the dilution of ferroelectric KNO by paraelectric KTO. Moreover, the standard deviation of the polarization magnitude increases notably with higher KTO content, indicating greater variability in local polarization. These observations can be attributed to the combined effects of lattice and polarization mismatches. The variation of average polarization magnitude is likely mainly associated with the elastic mismatch, where those of the KNO and KTO are suppressed and promoted, respectively. On the other hand, the increased standard deviation in polarization as a function of KTO content likely results from the electrostatic mismatch, where local polarization mismatch induces an internal electric field that disrupts the uniformity of polarization near interfaces. Together, these results further reinforce the critical roles of elastic and electrostatic mismatches in governing the structural and functional behavior of KNTO solid solutions, particularly in regions near PPBs.

2.5 Relaxor-like behavior in KNTO with high KTO concentrations

Relaxors are usually characterized by the formation of PNRs within a nonpolar matrix. The DP-simulated phase diagram (Figure 2(a)) shows that the coexistence of polar and nonpolar phases can be observed in KNTOs with KTO concentration between 40 mol % and 90 mol % near the Curie temperature. A defining feature of relaxors is their ability to undergo a field-induced transition into a long-range ferroelectric state when subjected to a sufficiently strong external electric field. To explore this, KNTO with 70 mol % KTO is simulated at 140 K under an applied electric field.

Figure 6 demonstrates the relaxor-like behavior of KNTO with 70 mol % KTO. In the absence of an electric field (Figure 6(a)), the system predominantly exhibits a non-polar phase, with a small fraction of polar phases. Upon application of a 20 kV/mm electric field, the polar phase fraction increases significantly, and importantly, this effect is reversible upon field removal, highlighting the ergodic nature of the relaxor. When subjected to an electric field of 20 kV/mm, the polar phase fraction significantly increases and is reversible upon field removal. The variation of phase fraction between the polar and non-polar phases should explain the reported enhancement of the electro-optical coefficient. The evolution of polarization during this process is shown in Figures 6(c)-6(e) and reveals a strong dependence on the local chemical composition, as mapped in Figure 6(b). Specifically, regions near KNO-rich sites exhibit stronger polarization than those dominated by KTO, consistent with the discussion in Section 2.4. A comparison is made for the same system simulated at 300K (pure non-polar phase) and 50K (pure polar phase), as shown in Figure S3.

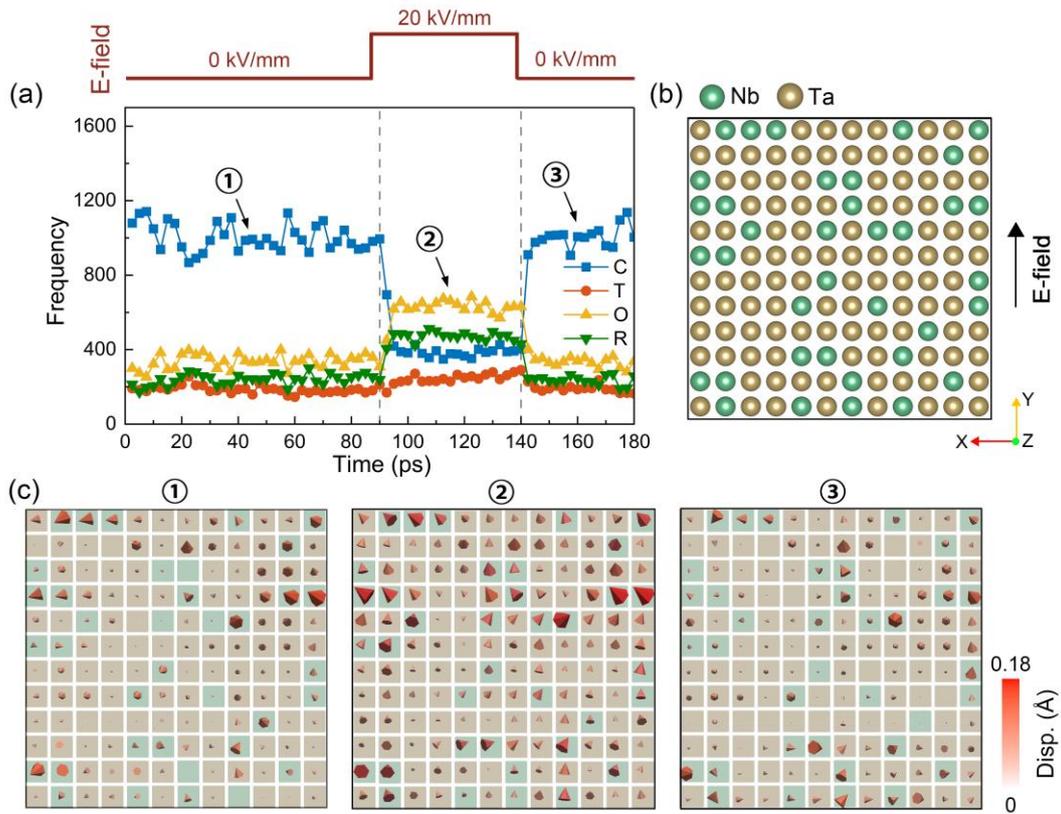

Figure 6. (a) Variation of phase structure at local unit cells in KNTO with 70 mol % KTO concentrations. (b) Chemical distribution of Nb and Ta in a 2-dimensional 12x12 region of interest sliced from the simulated 12x12x12 supercell. (c) Effective polarizations at the region of interest under different conditions, including state 1 (virgin), state 2 (applied electric field), and state 3 (after electric field removal). The cones represent the effective polarizations, while the squares in the background with green and brown colours correspond to the chemical distribution of Nb and Ta, respectively.

Furthermore, it is worth noting that the induced ferroelectric phases are mostly orthorhombic and rhombohedral phases, instead of the tetragonal phase, even though the electric field direction is strictly along [010] directions (one out of six possible polarization directions in the tetragonal phase). This is likely related to the energetic preference of orthorhombic and rhombohedral ferroelectric phases at a high KTO concentration (see Figure 2(c)).

Previous experimental studies have confirmed the relaxor nature of KNTOs at high KTO concentrations. The present DP-based simulations are in strong agreement with these findings and offer new insights into the atomic-scale origins of relaxor behavior, including the spatial distribution and evolution of PNRs under external fields. Future work could focus on further quantifying the size and dynamics of PNRs, their dependence on chemical homogeneity, and their influence on macroscopic functional properties.

## 3. Conclusion

In this work, we employed machine-learning molecular dynamics simulations to resolve the complex microstructures of the PPBs in KNTO solid solutions. We reveal that chemical composition and ordering critically govern the PPBs through elastic and electrostatic mismatches between ferroelectric $KNbO_3$ and paraelectric $KTaO_3$ components. Elastic strain is distributed homogeneously, while electrostatic mismatch induces localized disruptions near interfaces, leading to strongly diffused PPBs in chemically inhomogeneous systems. Moreover, we identify the emergence of relaxor-like behavior at high KTO concentrations, characterized by PNRs within a non-polar matrix and reversible field-induced transitions. These findings settle the longstanding debate over PPB microstructures and establish a generalizable framework for phase boundary engineering in complex solid solutions, paving the way for the practical design of next-generation high-performance ferroelectrics.

## 4. Methods

### 4.1 The DP model of KNTO

This study employed the open-source software DPGEN to construct a deep potential (DP) model for KNTO (*55, 56*). The workflow consists of three iterative steps: training, exploration, and labeling. First, an initial dataset was generated based on density functional theory (DFT) calculations, incorporating previously reported data for KNO. Supercells containing 40 atoms of KNTO in cubic, tetragonal, orthorhombic, and rhombohedral phases and 40-atom cubic KTO supercells were constructed, and DFT calculations were performed near the ground-state energy. The computational parameters were set as follows: within the VASP software package, the generalized gradient approximation (GGA) with the PBE functional was adopted, with a plane-wave cutoff energy of 900 eV, a k-point mesh spacing of 0.12 Å$^{-1}$, and convergence criteria of 10$^{-6}$ eV for energy and 0.005 eV/Å for forces. The DeePot-SE descriptor was utilized (*57*), with a neural network architecture comprising three embedding layers (with 20, 40, and 80 nodes, respectively) and three fitting layers (each containing 120 nodes). To accurately capture atomic interactions in the local environment, a cutoff radius of 9 Å was set for neighbor atom searches, along with a smooth cutoff of 1.5 Å.

Molecular dynamics (MD) simulation based on Large-scale Atomic/Molecular Massively Parallel Simulator (LAMMPS) was employed to explore the configurations across a temperature range of 300–900 K and a pressure range of 1–10000 bar, using the NPT ensemble with Nosé-Hoover thermostat. Poorly known configurations among models were identified based on force deviations within the range of 0.10-0.60 eV/Å. These configurations were subsequently subjected to labelling using DFT of the same standard as aforementioned. The iterative process continuously incorporated newly labeled data into the training dataset to refine the model. The final trained DP model was rigorously tested.

**4.2 MD simulation of KNTO**

4.2.1 Simulation of temperature-dependent phase transitions

The trained DP model was employed to perform large-scale molecular dynamics simulations using the LAMMPS software (*58*). A KNTO supercell containing 7980 atoms (12×12×12) was constructed using the Atomsk tool (*59*), and Ta was uniformly doped into the supercell, which was implemented using ASE in Python.

Firstly, the system was equilibrated for 10 ps in the NPT ensemble with the Nosé-Hoover thermostat at 805 K, with a timestep of 1 fs. Subsequently, the system was cooled from 805 K to 5 K at a rate of 10 K per 5 ps. During the simulations, time-averaged trajectories were recorded every 2.5 ps. Ultimately, the quantities of each phase structure within the supercell were calculated and partly visualized using Python, including ASE, Numpy, and Matplotlib. The effective polarizations are calculated by the relative position of the B-site atoms with respect to the center of the oxygen octahedra.

$$\vec{d} = \vec{B} - (\vec{O_1} + \vec{O_2} + \vec{O_3} + \vec{O_4} + \vec{O_5} + \vec{O_6})/6$$

Here, $\vec{B}$ represents the coordinate vector of the B-site atoms, and $\vec{O_i}$ denote the coordinate vectors of the six neighboring oxygen atoms within the oxygen octahedron.

4.2.2 Simulation of electric field application

To simulate the electric field, Born effective charges of KNTO (*60*) is obtained from DFT calculation and set as fixed charges, which are as follows: $Z_K^* = 1.143, Z_{Nb}^* = 9.123$, $Z_{Ta}^* = 9.323$, and $Z_O^* = -3.410$. For the KNTO system with a Ta concentration of 0.70, a multi-stage relaxation process was carried out. Initially, the system was relaxed for 20 ps in the NVT ensemble with the Nosé-Hoover thermostat, followed by 90 ps in the NPT ensemble with the Nosé-Hoover thermostat to achieve an equilibrium state at the target temperature and pressure. Then, an external electric field of 0.004 V/Å along the y-direction was applied under the NVT ensemble with Langevin thermostat for 50 ps to simulate the effect of the electric field on the system. Finally, the system was relaxed for 40 ps in the NPT ensemble without the electric field. The results are jointly visualized using VESTA (*61*) and Paraview (*62*).

### 4.3 The phonon spectra

In the supplementary materials, the phonon spectra (Figure S1) are calculated using the finite displacement method, and the non-analytical term correction is implemented using PHONOPY (*63*). A 3x3x3 supercell is created for phonon calculation. The software automatically estimates band connections with 101 sampling points along each path.

### 4.4 Experimental preparation and characterization of KNTO ceramics

4.4.1 Ceramic Processing

$KNb_{1-x}Ta_xO_3$ ($x$ = 0, 0.1, 0.2, 0.3) powder was prepared via solid-state reaction. Raw materials $K_2CO_3$ (99.8%, Sinopharm), $Ta_2O_5$ (99.99%, Sinopharm), and $Nb_2O_5$ (99.99%, Sinopharm) were dried at 120 °C for 24 h and subsequently weighed in a stoichiometric ratio. All raw materials were homogenized by planetary ball milling using yttria-stabilized zirconia balls in ethanol for 24 h at 300 rpm and then dried at 120 °C. The powder mixtures were first calcined in covered alumina crucibles at 960 °C for 4 h. The as-calcined powders were subsequently ball-milled for 24 h at 300 rpm, dried at 120 °C. After that, the powders were calcined again at 1000 °C for 4 h. Then, the twice-calcined powders were ball-milled again for 24 h at 300 rpm and dried at 120 °C. Afterwards, 0.25g of the calcined powders were pressed into 10-mm diameter discs and set using cold isostatic pressing. Finally, the green bodies are sintered in a muffle furnace. Four ceramics (*i.e.*, $KNbO_3$, $KNb_{0.9}Ta_{0.1}O_3$, $KNb_{0.8}Ta_{0.2}O_3$, $KNb_{0.7}Ta_{0.3}O_3$) were sintered at 1030°C, 1050°C, 1060°C, and 1080°C for 4 hours, respectively.

4.4.2 Characterizations

The microstructure of the ceramics was probed by using a scanning electron microscope (SEM, JSM-6460LV, JEOL, Japan). The ceramics were polished and thermally etched for microstructure observation. The elemental distribution of the samples (polished but not thermally etched) was probed by Electron Probe X-ray Micro-Analyzer (EPMA, JXA8230, JEOL, Japan).

4.4.3 Characterization of Electrical Properties

For macroscopic electrical property characterizations, the ceramics were painted with silver pastes and fired at 600 °C for 30 min to form electrodes. The temperature-dependent dielectric constant from room temperature to 500 °C and the impedance spectra at a frequency of 100 kHz.

## 5. Acknowledgement

This work was supported by the National Natural Science Foundation of China (Nos. 52325204, 52421001, W2433118 and U2241243) and the Beijing Natural Science Foundation (Nos. IS24026 and JQ22010).

## 6. Declaration

The authors declare that they have no competing interests.

## 7. Data availability

All data needed to evaluate the conclusions in the paper are present in the paper and/or the Supplementary Materials.

Supplementary Materials for

**Machine learning-enabled atomistic insights into phase boundary engineering of solid-solution ferroelectrics**

Weiru Wen[1,2], Fan-Da Zeng[2], Ben Xu[3], Bi Ke[1,*], Zhipeng Xing[2,*], Hao-Cheng Thong[2,*], Ke Wang[2]

[1] State Key Laboratory of Information Photonics and Optical Communications, School of Physical Science and Technology, Beijing University of Posts and Telecommunications, Beijing, 100876, China

[2] State Key Laboratory of New Ceramic Materials, School of Materials Science and Engineering, Tsinghua University, Beijing 100084, P. R. China

[3] Graduate School, China Academy of Engineering Physics, Beijing 100193, People's Republic of China

* Corresponding authors: Bi Ke (bike@bupt.edu.cn); Zhipeng Xing (xingzhipeng@tsinghua.edu.cn); Hao-Cheng Thong (haocheng@mail.tsinghua.edu.cn)

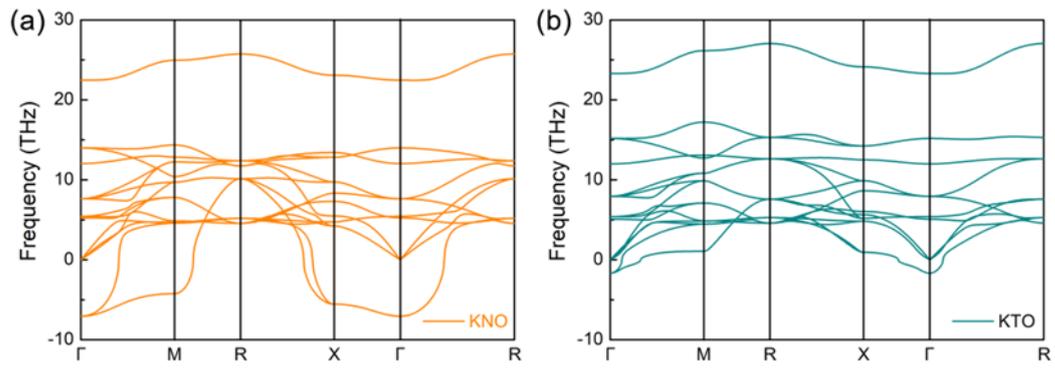

Figure S1. Phonon dispersion spectra of cubic (a) KNO and (b) KTO.

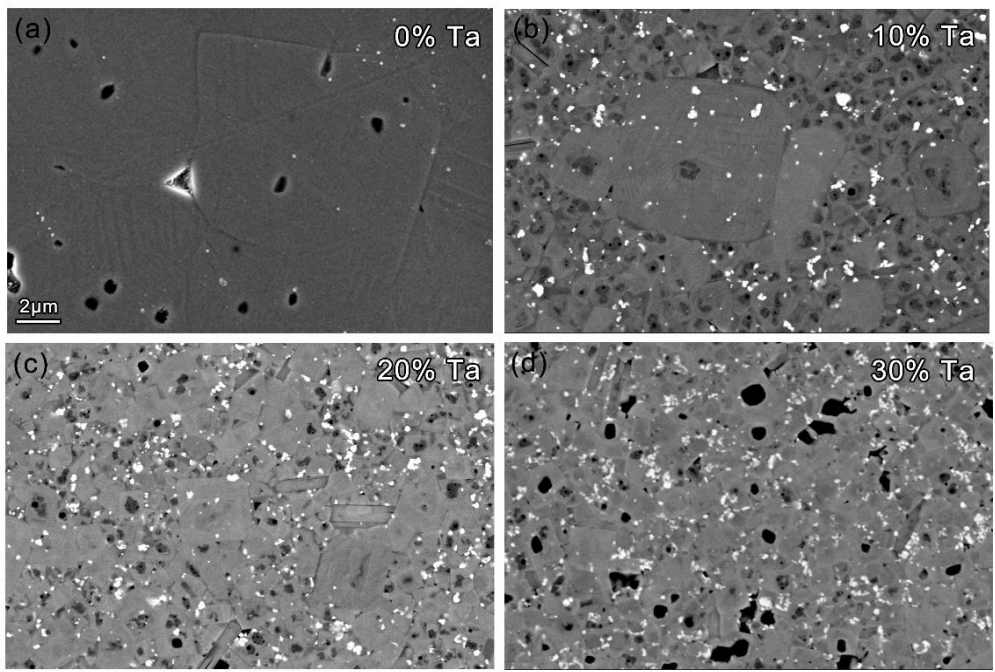

Figure S2. SEM back-scattered electron (BSE) image of as-sintered K(Nb,Ta)O$_3$ sample (a)-(d) with Ta concentrations of 0%, 10%, 20%, and 30%, respectively. The scale bar is 2 μm.

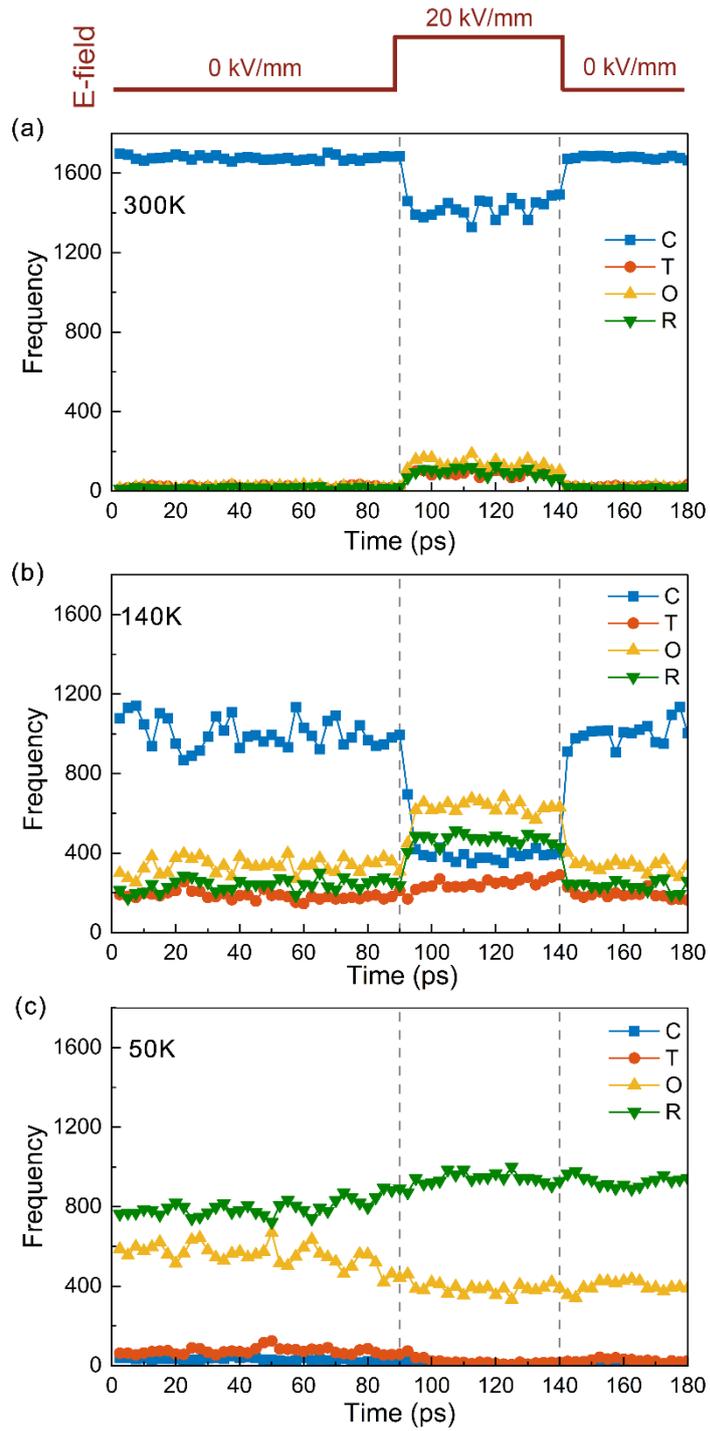

Figure S3. Variation of phase structure at local unit cells in KNTO with 70 mol% KTO concentrations at different temperatures: (a) 300 K; (b) 140 K; (c) 50 K.